\documentclass[conference,romanappendices,10pt]{IEEEtran}

\IEEEoverridecommandlockouts

\usepackage{algorithm,algorithmic,amsmath,amssymb,amsthm,bbm,caption,cite,color,comment,graphicx,microtype,multicol,setspace,subcaption,tabularx,url}
\usepackage[USenglish]{babel}
\usepackage[T1]{fontenc}
\usepackage[utf8]{inputenc}

\usepackage{enumitem}
\setitemize{itemsep=0mm,topsep=0mm,parsep=0pt,partopsep=0pt}

\makeatletter
\newcommand\subparagraph{%
  \@startsection{subparagraph}{5}
  {\parindent}
  {3.25ex \@plus 1ex \@minus .2ex}
  {-1em}
  {\normalfont\normalsize\bfseries}}
\makeatother
\usepackage{titlesec}
\let\subparagraph\relax

\usepackage{titlesec}
\let\subparagraph\relax
\titlespacing{\section}{0pt}{5pt plus 2pt minus 1pt}{3pt plus 1pt minus 0pt}
\titlespacing{\subsection}{0pt}{4pt plus 2pt minus 1pt}{2pt plus 1pt minus 0pt}
\usepackage[figurename=Fig.,font=small]{caption}

{}
{}
{}
{}
{}
{}
{}

\renewcommand{\d}{\mathbf{d}}

\newcommand{\h}{\mathbf{h}}

\newcommand{\p}{\mathbf{p}}

\renewcommand{\r}{\mathbf{r}}

\renewcommand{\v}{\mathbf{v}}
\newcommand{\w}{\mathbf{w}}
\newcommand{\x}{\mathbf{x}}
\newcommand{\y}{\mathbf{y}}
\newcommand{\z}{\mathbf{z}}

\newcommand{\0}{\mathbf{0}}

\newcommand{\C}{\mathbf{C}}

\newcommand{\G}{\mathbf{G}}
\renewcommand{\H}{\mathbf{H}}
\newcommand{\I}{\mathbf{I}}

\renewcommand{\P}{\mathbf{P}}

\newcommand{\V}{\mathbf{V}}
\newcommand{\W}{\mathbf{W}}

\newcommand{\setC}{\mathcal{C}}

\newcommand{\setL}{\mathcal{L}}

\newcommand{\setN}{\mathcal{N}}

\newcommand{\setT}{\mathcal{T}}

\newcommand{\Compl}{\mbox{$\mathbb{C}$}}
\newcommand{\rmF}{\mathrm{F}}

\newcommand{\Diag}{\mathrm{Diag}}

\newcommand{\Exp}{\mathbb{E}}
\newcommand{\herm}{\mathrm{H}}
\renewcommand{\Im}{\mathrm{Im}}

\renewcommand{\Re}{\mathrm{Re}}

\newcommand{\tr}{\mathrm{tr}}
\newcommand{\tran}{\mathrm{T}}

\newcommand{\adc}{\textrm{\tiny{ADC}}}
\newcommand{\bs}{\textrm{\tiny{BS}}}
\newcommand{\ce}{\textrm{\tiny{ce}}}
\newcommand{\dac}{\textrm{\tiny{DAC}}}
\newcommand{\dl}{\textrm{\tiny{dl}}}
\newcommand{\hw}{\textrm{\tiny{HW}}}

\newcommand{\rf}{\textrm{\tiny{RF}}}

\newcommand{\ue}{\textrm{\tiny{UE}}}
\newcommand{\ul}{\textrm{\tiny{ul}}}

\usepackage{fancyhdr}
\fancypagestyle{firstpage}{
  
  \fancyhead[R]{{\scriptsize 1}}
  \fancyfoot[C]{{\footnotesize \begin{singlespace} \textcopyright 2021 IEEE. Personal use of this material is permitted. Permission from IEEE must be obtained for all other uses, in any current or future media, including reprinting/republishing this material for advertising or promotional purposes, creating new collective works, for resale or redistribution to servers or lists, or reuse of any copyrighted component of this work in other works. \end{singlespace}}}}

\title{Low-Resolution Massive MIMO Under \\ Hardware Power Consumption Constraints}
\author{
\IEEEauthorblockN{Italo Atzeni,$^{1}$ Antti Tölli,$^{1}$ and Giuseppe Durisi$^{2}$} \\ 
\IEEEauthorblockA{$^{1}$Centre for Wireless Communications, University of Oulu, Finland \\
$^{2}$Chalmers University of Technology, Sweden \\
Emails: \{italo.atzeni, antti.tolli\}@oulu.fi, durisi@chalmers.se}
\thanks{\vspace{-2mm}

The work of I.~Atzeni was supported by the Marie Sk\l{}odowska-Curie Actions (MSCA-IF 897938 DELIGHT). The work of A.~Tölli was supported by the Academy of Finland (318927 6Genesis Flagship and 319059 CCCWEE). The work of G.~Durisi was supported by the Wallenberg AI, Autonomous Systems, and Software Program.} \vspace{-5mm}}

\begin{document}

\maketitle

\thispagestyle{firstpage}

\begin{abstract}
We consider a fully digital massive multiple-input multiple-output architecture with low-resolution analog-to-digital/digital-to-analog converters (ADCs/DACs) at the base station (BS) and analyze the performance trade-off between the number of BS antennas, the resolution of the ADCs/DACs, and the bandwidth. Assuming a hardware power consumption constraint, we determine the relationship between these design parameters by using a realistic model for the power consumption of the ADCs/DACs and the radio frequency chains. Considering uplink pilot-aided channel estimation, we build on the Bussgang decomposition to derive tractable expressions for uplink and downlink ergodic achievable sum rates. Numerical results show that the ergodic performance is boosted when many BS antennas with very low resolution (i.e., $2$ to $3$ bits) are adopted in both the uplink and the downlink.
\end{abstract}

\section{Introduction} \label{sec:Intro}

To support the exponential growth in mobile data traffic, beyond-5G wireless systems are expected to exploit the large amount of bandwidth available in the sub-THz band (0.1--1 THz) \cite{Raj20}. In this context, fully digital architectures allow one to realize the full potential of massive multiple-input multiple-output (MIMO) by enabling highly flexible beamforming and large-scale spatial multiplexing. In fully digital massive MIMO, each base station (BS) antenna is connected to a radio-frequency (RF) chain that includes a pair of power-hungry analog-to-digital/digital-to-analog converters (ADCs/DACs). In this setting, the power consumption of each ADC/DAC scales linearly with the sampling rate and exponentially with the number of resolution bits \cite{Jac17,Li17}. Hence, adopting low-resolution ADCs/DACs (even down to $1$-bit \cite{Atz22,Atz21}) represents a necessary step towards the energy-efficient implementation of fully digital massive MIMO arrays comprising hundreds or thousands of antennas.

In this paper, we analyze the performance trade-off between the number of BS antennas, the resolution of the ADCs/DACs, and the bandwidth under a hardware power consumption constraint. The relationship between these design parameters is determined by means of a realistic model for the power consumption of the ADCs/DACs and the RF chains. In this setting, a lower resolution of the ADCs/DACs allows one to increase the number of BS antennas (and, consequently, the array gain and the spatial resolution) at the expense of a more significant quantization distortion during both the channel estimation and the uplink/downlink data transmission. To analyze this trade-off, we derive tractable expressions for uplink and downlink ergodic achievable sum rates with imperfect channel state information (CSI). Leveraging the Bussgang decomposition \cite{Bus52}, we take into account the quantization distortion in the uplink/downlink data transmission as well as in the combining and precoding matrices, which are based on quantized channel estimates. To further simplify the resulting sum rate expressions, we assume that the BS adopts maximum ratio combining (MRC) in the uplink and maximum ratio transmission (MRT) in the downlink. In the evaluated scenarios, the most energy-efficient configurations consist of many antennas with $2$ to $3$ resolution bits in both the uplink and the downlink. A related study was presented in \cite{Jac18,Ett19}, where the uplink and downlink outage rates were analyzed in the presence of a fronthaul constraint.

\section{System Model} \label{sec:SM}

We consider a massive MIMO system where a BS with $M$ antennas and low-resolution ADCs/DACs serves $K$ single-antenna user equipments (UEs) with infinite-resolution ADCs/DACs. We use $\H \triangleq [\h_{1}, \ldots, \h_{K}] \in \Compl^{M \times K}$ to denote the uplink channel matrix.\footnote{We will consider the special case of i.i.d. Rayleigh fading channels in Section~\ref{sec:SR_imperf}.} Each BS antenna is connected to two $b$-bit ADCs in the uplink (resp. to two $b$-bit DACs in the downlink), one for the in-phase and one for the quadrature component of the receive (resp. transmit) signal. Each ADC/DAC in the uplink/downlink consists of a $b$-bit quantizer characterized by a set of $2^{b} + 1$ quantization thresholds $\setT_{b} \triangleq \{ t_{n} \}_{n=0}^{2^{b}}$, with $- \infty = t_{0} < \ldots < t_{2^{b}} = \infty$, and by a set of $2^{b}$ quantization labels $\setL_{b} \triangleq \{ \ell_{n} \}_{n=0}^{2^{b}-1}$, with $\ell_{n} \in (t_{n}, t_{n+1}]$. In this setting, we introduce the quantization function $Q_{b} (\cdot)$ such that, for $s \in \Compl$, $Q_{b} (s) = \ell_{i} + j \, \ell_{k}$ if $\Re[s] \in (t_{i}, t_{i+1}]$ and $\Im[s] \in (t_{k}, t_{k+1}]$; for a matrix or vector input, $Q_{b} (\cdot)$ is applied entrywise.

We assume that the number of BS antennas, the resolution of the ADCs/DACs, and the bandwidth are related through a hardware power consumption constraint. Let $P_{\hw}$ be the maximum hardware power consumption allowed at the BS and let $P_{\adc}(b,B)$ and $P_{\dac}(b,B)$ denote the power consumed by each ADC and DAC, respectively, where $B$ represents the bandwidth. Both $P_{\adc}(b,B)$ and $P_{\dac}(b,B)$ grow linearly with the bandwidth and exponentially with the number of resolution bits, as in the power consumption model adopted in Section~\ref{sec:NR_quant}. Hence, the number of BS antennas available in the uplink/downlink for a given resolution of the ADCs/DACs $b$ and a given bandwidth $B$ is
\begin{align} \label{eq:M}
M = \begin{cases}
\Bigl\lfloor \frac{P_{\hw}}{P_{\rf} + 2 P_{\adc}(b,B)} \Bigr\rfloor & \textrm{(uplink)}, \vspace{1mm} \\
\Bigl\lfloor \frac{P_{\hw}}{P_{\rf} + 2 P_{\dac}(b,B)} \Bigr\rfloor & \textrm{(downlink)}
\end{cases}
\end{align}
where $P_{\rf}$ denotes the power consumed by each RF chain. Note that, in the following, $M$ assumes different values for the uplink and the downlink according to \eqref{eq:M}.

\begin{figure*}
\addtocounter{equation}{+8}
\begin{align}
\label{eq:gamma^ul} \gamma_{k}^{\ul} & \triangleq \frac{\rho_{\bs} \big| \Exp [\v_{k}^{\herm} \G_{b}^{\ul} \h_{k}] \big|^{2}}{\rho_{\bs} \sum_{i=1}^{K} \Exp \big[ |\v_{k}^{\herm} \G_{b}^{\ul} \h_{i}|^{2} \big] - \rho_{\bs} \big| \Exp [\v_{k}^{\herm} \G_{b}^{\ul} \h_{k}] \big|^{2} + \Exp \big[ \| \G_{b}^{\ul} \v_{k} \|^{2} \big] + \Exp [\v_{k}^{\herm} \C_{\d}^{\ul} \v_{k}]}, \\ \addtocounter{equation}{+7}
\label{eq:gamma^dl} \gamma_{k}^{\dl} & \triangleq \frac{\rho_{\ue} \big| \Exp [\h_{k}^{\herm} \G_{b}^{\dl} \w_{k}] \big|^{2}}{\rho_{\ue} \sum_{i=1}^{K} \Exp \big[ |\h_{k}^{\herm} \G_{b}^{\dl} \w_{i}|^{2} \big] - \rho_{\ue} \big| \Exp [\h_{k}^{\herm} \G_{b}^{\dl} \w_{k}] \big|^{2} + \rho_{\ue} \Exp [\h_{k}^{\herm} \C_{\d}^{\dl} \h_{k}] + 1}
\end{align}
\hrulefill
\vspace{-5mm}
\end{figure*}

\section{Ergodic Performance} \label{sec:SR_gen}

In this section, we provide expressions for uplink and downlink ergodic achievable sum rates with $b$-bit ADCs and DACs, respectively, for general combining and precoding matrices. These are extended in Section~\ref{sec:SR_imperf} to include the effect of imperfect CSI arising from the uplink pilot-aided channel estimation with low-resolution ADCs.

\subsection{Uplink Ergodic Achievable Sum Rate} \label{sec:SR_gen_UL}

Let $\x^{\ul} \sim \setC \setN (\0, \I_{K})$ denote the uplink transmit symbol vector. At the BS, the normalized input signal at the ADCs is given by
\addtocounter{equation}{-17}
\begin{align} \label{eq:y^ul}
\y^{\ul} \triangleq \sqrt{\rho_{\bs}} \H \x^{\ul} + \z^{\ul} \in \Compl^{M \times 1}
\end{align}
where $\z^{\ul} \sim \setC \setN (\0, \I_{M})$ is the additive white Gaussian noise (AWGN) vector; since the AWGN power is normalized to one, $\rho_{\bs}$ represents the signal-to-noise ratio (SNR) at the BS. Then, at the output of the ADCs, we have
\begin{align} \label{eq:r^ul}
\r^{\ul} \triangleq Q_{b} (\y^{\ul}) \in \Compl^{M \times 1}.
\end{align}
We assume that the BS obtains a soft estimate of $\x^{\ul}$ as
\begin{align}
\hat{\x}^{\ul} & \triangleq [\hat{x}_{1}^{\ul}, \ldots, \hat{x}_{K}^{\ul}]^{\tran} \\
\label{eq:x_hat^ul} & = \V^{\herm} \r^{\ul} \in \Compl^{K \times 1}
\end{align}
where $\V \triangleq [\v_{1}, \ldots, \v_{K}] \in \Compl^{M \times K}$ is the combining matrix.

Using the Bussgang decomposition, we can write \eqref{eq:r^ul} as
\begin{align} \label{eq:r^ul_2}
\r^{\ul} & = \G_{b}^{\ul} \y^{\ul} + \d^{\ul}
\end{align}
where $\d^{\ul} \in \Compl^{M \times 1}$ is the zero-mean, non-Gaussian quantization distortion vector that is uncorrelated with $\y^{\ul}$ and has covariance matrix $\C_{\d}^{\ul} \triangleq \Exp \big[ \d^{\ul} (\d^{\ul})^{\herm} \big] \in \Compl^{M \times M}$. In addition, $\G_{b}^{\ul} \in \Compl^{M \times M}$ is a diagonal matrix given by (see, e.g., \cite{Jac17})
\begin{align}
\nonumber \G_{b}^{\ul} & \triangleq \big( \pi \Diag (\C_{\y}^{\ul}) \big)^{-\frac{1}{2}} \sum_{n=0}^{2^{b}-1} \ell_{n} \Big( \exp \big( - t_{n}^{2} \Diag(\C_{\y}^{\ul})^{-1} \big) \\
\label{eq:G_b_ul} & \phantom{=} \ - \exp \big( - t_{n+1}^{2} \Diag(\C_{\y}^{\ul})^{-1} \big) \Big)
\end{align}
with $\C_{\y}^{\ul} \triangleq \Exp \big[ \y^{\ul} (\y^{\ul})^{\herm} \big] \in \Compl^{M \times M}$. Substituting \eqref{eq:r^ul_2} into \eqref{eq:x_hat^ul}, we obtain
\begin{align} \label{eq:x_hat^ul_2}
\hat{\x}^{\ul} = \V^{\herm} \big( \G_{b}^{\ul} (\sqrt{\rho_{\bs}} \H \x^{\ul} + \z^{\ul}) + \d^{\ul} \big).
\end{align}
To derive an easy-to-evaluate expression for an uplink ergodic achievable sum rate, we use the use-and-then-forget bound \linebreak \cite[Thm.~4.4]{Bjo17}. Specifically, we write the $k$th entry of \eqref{eq:x_hat^ul_2} as

\newpage

$ $ \vspace{-8.5mm}

\begin{align}
\nonumber \hat{x}_{k}^{\ul} & = \sqrt{\rho_{\bs}} \Exp [\v_{k}^{\herm} \G_{b}^{\ul} \h_{k}] x_{k}^{\ul} \! + \! \sqrt{\rho_{\bs}} \big( \v_{k}^{\herm} \G_{b}^{\ul} \h_{k} \! - \! \Exp [\v_{k}^{\herm} \G_{b}^{\ul} \h_{k}] \big) x_{k}^{\ul} \\
& \phantom{=} \ + \sqrt{\rho_{\bs}} \sum_{i \neq k} \v_{k}^{\herm} \G_{b}^{\ul} \h_{i} x_{i}^{\ul} + \v_{k}^{\herm} \G_{b}^{\ul} \z^{\ul} + \v_{k}^{\herm} \d^{\ul}.
\end{align}
Treating $\d^{\ul}$ as Gaussian noise with covariance matrix $\C_{\d}^{\ul}$ and assuming that the BS is aware of the value of $\Exp [\v_{k}^{\herm} \G_{b}^{\ul} \h_{k}]$, we obtain the effective uplink signal-to-interference-plus-noise-and-distortion ratio (SINDR) of UE~$k$ as in \eqref{eq:gamma^ul} at the top of the next page. Finally, an uplink ergodic achievable sum rate is given by \vspace{-1mm}
\addtocounter{equation}{+1}
\begin{align} \label{eq:R^ul}
R^{\ul} \triangleq B \sum_{k=1}^{K} \log_{2}(1 + \gamma_{k}^{\ul}).
\end{align}

\subsection{Downlink Ergodic Achievable Sum Rate} \label{sec:SR_gen_DL}

Let $\x^{\dl} \sim \setC \setN (\0, \I_{K})$ denote the downlink transmit symbol vector. At the BS, the output signal at the DACs is given by
\begin{align} \label{eq:r^dl}
\r^{\dl} \triangleq Q_{b} (\W \x^{\dl}) \in \Compl^{M \times 1}
\end{align}
where $\W \triangleq [\w_{1}, \ldots, \w_{K}] \in \Compl^{M \times K}$ is the precoding matrix with $\Exp \big[ \| \W \|_{\rmF}^{2}] = 1$. Then, the normalized receive signal at the UEs is given by
\begin{align}
\y^{\dl} & \triangleq [y_{1}^{\dl}, \ldots, y_{K}^{\dl}]^{\tran} \\
\label{eq:y^dl} & = \sqrt{\rho_{\ue}} \H^{\herm} \r^{\dl} + \z^{\dl} \in \Compl^{K \times 1}
\end{align}
where $\z^{\dl} \sim \setC \setN (\0, \I_{K})$ is the AWGN vector; since the AWGN power is normalized to one, $\rho_{\ue}$ represents the SNR at the UEs.

Using the Bussgang decomposition, we can write \eqref{eq:r^dl} as
\begin{align} \label{eq:r^dl_2}
\r^{\dl} = \G_{b}^{\dl} \W \x^{\dl} + \d^{\dl}
\end{align}
where $\d^{\dl} \in \Compl^{M \times 1}$ is the zero-mean, non-Gaussian quantization distortion vector that is uncorrelated with $\W \x^{\dl}$ and has covariance matrix $\C_{\d}^{\dl} \triangleq \Exp \big[ \d^{\dl} (\d^{\dl})^{\herm} \big] \in \Compl^{M \times M}$. In addition, $\G_{b}^{\dl} \in \Compl^{M \times M}$ is a diagonal matrix obtained by replacing $\C_{\y}^{\ul}$ in \eqref{eq:G_b_ul} with $\Exp [\W \W^{\herm}]$. Substituting \eqref{eq:r^dl_2} into \eqref{eq:y^dl}, we obtain
\begin{align} \label{eq:y^dl_2}
\y^{\dl} = \sqrt{\rho_{\ue}} \H^{\herm} (\G_{b}^{\dl} \W \x^{\dl} + \d^{\dl}) + \z^{\dl}.
\end{align}
To derive an easy-to-evaluate expression for a downlink ergodic achievable sum rate, we use the channel hardening bound \cite[Thm.~4.6]{Bjo17}. Specifically, we write the $k$th entry of \eqref{eq:y^dl_2} as
\begin{align}
\nonumber y_{k}^{\dl} & \! = \! \sqrt{\rho_{\ue}} \Exp [\h_{k}^{\herm} \G_{b}^{\dl} \w_{k}] x_{k}^{\dl} \! + \! \sqrt{\rho_{\ue}} \big( \h_{k}^{\herm} \G_{b}^{\dl} \w_{k} \! - \! \Exp [\h_{k}^{\herm} \G_{b}^{\dl} \w_{k}] \big) x_{k}^{\dl} \\
& \phantom{=} \ \! + \sqrt{\rho_{\ue}} \sum_{i \neq k} \h_{k}^{\herm} \G_{b}^{\dl} \w_{i} x_{i}^{\dl} + \sqrt{\rho_{\ue}} \h_{k}^{\herm} \d^{\dl} + z_{k}^{\dl}.
\end{align}
Treating $\d^{\dl}$ as Gaussian noise with covariance matrix $\C_{\d}^{\dl}$ and assuming that UE~$k$ is aware of the value of $\Exp [\h_{k}^{\herm} \G_{b}^{\dl} \w_{k}]$, we obtain the effective downlink SINDR of UE~$k$ as in \eqref{eq:gamma^dl} at the top of the page. Finally, a downlink ergodic achievable sum rate is given by \vspace{-1mm}
\addtocounter{equation}{+1}
\begin{align} \label{eq:R^dl}
R^{\dl} \triangleq B \sum_{k=1}^{K} \log_{2}(1 + \gamma_{k}^{\dl}).
\end{align}

\section{Ergodic Performance with Imperfect CSI} \label{sec:SR_imperf}

The ergodic achievable sum rate expressions derived in Section~\ref{sec:SR_gen} take into account the quantization distortion in the uplink/downlink data transmission. In practice, however, the quantization also affects the combining and precoding matrices, which are based on quantized channel estimates. In this section, we include the effect of imperfect CSI arising from the uplink pilot-aided channel estimation with low-resolution ADCs. To further simplify the resulting sum rate expressions, we consider the special case of i.i.d. Rayleigh fading channels, where the entries of $\H$ are independent $\setC \setN (0,1)$ random variables, and we assume that the BS adopts MRC in the uplink and MRT in the downlink. The full derivations are omitted due to the space limitations and will be included in an extended version of this paper, which will also consider correlated channels.

\begin{figure*}
\addtocounter{equation}{+13}
\begin{align}
\label{eq:gamma^ul_2} \gamma_{k}^{\ul} & = \frac{\rho_{\bs} (G_{b}^{\ce})^{2} (G_{b}^{\ul})^{2} M^{2}}{(\rho_{\bs} K + 1) \big( 1 + \frac{1}{\rho_{\bs} \tau} \big) (G_{b}^{\ce})^{2} (G_{b}^{\ul})^{2} M + (\rho_{\bs} K + 1) \frac{1}{\rho_{\bs} \tau^{2}} (G_{b}^{\ul})^{2} A_{k} + \big( 1 + \frac{1}{\rho_{\bs} \tau} \big) (G_{b}^{\ce})^{2} \tr (\C_{\d}^{\ul}) + \frac{1}{\rho_{\bs} \tau^{2}} B_{k}}, \\ \addtocounter{equation}{+6}
\label{eq:gamma^dl_2} \gamma_{k}^{\dl} & = \frac{\rho_{\ue} (G_{b}^{\ce})^{2} (G_{b}^{\dl})^{2} M^{2}}{\rho_{\ue} K \big( 1 + \frac{1}{\rho_{\bs} \tau} \big) (G_{b}^{\ce})^{2} (G_{b}^{\dl})^{2} M + \rho_{\ue} \frac{1}{\rho_{\bs} \tau^{2}} (G_{b}^{\dl})^{2} \sum_{i=1}^{K} A_{i} + \delta \rho_{\ue} \tr (\C_{\d}^{\dl}) + \delta}
\end{align}
\hrulefill
\vspace{-5.5mm}
\end{figure*}

\subsection{Channel Estimation with $b$-Bit ADCs} \label{sec:SR_imperf_CE}

During the uplink pilot-aided channel estimation phase, the UEs simultaneously transmit their uplink pilots of length $\tau$. Let $\P \triangleq [\p_{1}, \ldots, \p_{K}] \in \Compl^{\tau \times K}$ denote the pilot matrix, where $\p_{k} \in \Compl^{\tau \times 1}$ represents the pilot sequence assigned to UE~$k$. In this context, we assume $\tau \geq K$ and that the pilot sequences are orthogonal, so that $\P^{\herm} \P = \tau \I_{K}$. Furthermore, we define $\bar{\P}_{k} \triangleq \p_{k} \otimes \I_{M} \in \Compl^{M \tau \times M}$ and $\bar{\P} \triangleq [\bar{\P}_{1}, \ldots, \bar{\P}_{K}] \in \Compl^{M \tau \times M K}$. At the BS, the normalized input signal at the ADCs is given by
\addtocounter{equation}{-21}
\begin{align} \label{eq:y^ce}
\y^{\ce} \triangleq \sqrt{\rho_{\bs}} \bar{\P}^{*} \h + \z^{\ce} \in \Compl^{M \tau \times 1}
\end{align}
with $\h \triangleq \mathrm{vec} (\H) \in \Compl^{M K \times 1}$ and where $\z^{\ce} \sim \setC \setN (\0, \I_{M \tau})$ is the AWGN vector. Then, at the output of the ADCs, we have
\begin{align} \label{eq:r^ce}
\r^{\ce} & \triangleq Q_{b} (\y^{\ce}) \in \Compl^{M \tau \times 1}.
\end{align}
We assume that the BS obtains an estimate of $\h$ (up to a scaling factor) as
\begin{align} \label{eq:h_hat}
\hat{\h} \triangleq \frac{1}{\sqrt{\rho_{\bs}} \tau} \bar{\P}^{\tran} \r^{\ce} \in \Compl^{M K \times 1}.
\end{align}

Using the Bussgang decomposition, we can write \eqref{eq:r^ce} as
\begin{align} \label{eq:r^ce_2}
\r^{\ce} & = \G_{b}^{\ce} \y^{\ce} + \d^{\ce}
\end{align}
where $\d^{\ce} \in \Compl^{M \tau \times 1}$ is the zero-mean, non-Gaussian quantization distortion vector that is uncorrelated with $\y^{\ce}$ and has covariance matrix $\C_{\d}^{\ce} \triangleq \Exp \big[ \d^{\ce} (\d^{\ce})^{\herm} \big] \in \Compl^{M \tau \times M \tau}$. In addition, $\G_{b}^{\ce} \in \Compl^{M \tau \times M \tau}$ is a diagonal matrix obtained by replacing $\C_{\y}^{\ul}$ in \eqref{eq:G_b_ul} with $\C_{\y}^{\ce} \triangleq \Exp \big[ \y^{\ce} (\y^{\ce})^{\herm} \big] \in \Compl^{M \tau \times M \tau}$. Substituting \eqref{eq:r^ce_2} into \eqref{eq:h_hat}, we obtain
\begin{align} \label{eq:h_hat_2}
\hat{\h} & = \frac{1}{\tau} \bigg( \bar{\P}^{\tran} \G_{b}^{\ce} \bar{\P}^{*} \h + \frac{1}{\sqrt{\rho_{\bs}}} \bar{\P}^{\tran} \G_{b}^{\ce} \z^{\ce} + \frac{1}{\sqrt{\rho_{\bs}}} \bar{\P}^{\tran} \d^{\ce} \bigg).
\end{align}

\subsection{Uplink Ergodic Achievable Sum Rate with MRC} \label{sec:SR_imperf_UL}

Assuming that MRC is used at the BS, we have $\V = \G_{b}^{\ul} \hat{\H}$ and the combining vector corresponding to UE~$k$ is given by
\begin{align}
\v_{k} & = \frac{1}{\tau} \G_{b}^{\ul} \bigg( \bar{\P}_{k}^{\tran} \G_{b}^{\ce} \bar{\P}^{*} \h + \frac{1}{\sqrt{\rho_{\bs}}} \bar{\P}_{k}^{\tran} \G_{b}^{\ce} \z^{\ce} + \frac{1}{\sqrt{\rho_{\bs}}} \bar{\P}_{k}^{\tran} \d^{\ce} \bigg).
\end{align}
Under the i.i.d. Rayleigh fading assumption, we have $\C_{\y}^{\ul} = (\rho_{\bs} K + 1) \I_{M}$ and $\C_{\y}^{\ce} = \rho_{\bs} \bar{\P}^{*} \bar{\P}^{\tran} + \I_{M \tau}$ (with $\Diag(\C_{\y}^{\ul}) = (\rho_{\bs} K + 1) \I_{M \tau}$), and we can thus write $\G_{b}^{\ul}$ and $\G_{b}^{\ce}$ as scaled identity matrices, i.e., $\G_{b}^{\ul} = G_{b}^{\ul} \I_{M}$ and $\G_{b}^{\ce} = G_{b}^{\ce} \I_{M \tau}$ (see \eqref{eq:G_b_ul}). Moreover, we introduce the following preliminary definitions:

\newpage

$ $ \vspace{-9mm}

\begin{align}
A_{k} & \triangleq \tr (\C_{\d}^{\ce} \bar{\P}_{k}^{*} \bar{\P}_{k}^{\tran}), \\
B_{k} & \triangleq \tr (\C_{\d}^{\ce} \bar{\P}_{k}^{*} \C_{\d}^{\ul} \bar{\P}_{k}^{\tran}).
\end{align}
Then, we can express each expectation term in \eqref{eq:gamma^ul} as
\begin{align}
\Exp [\v_{k}^{\herm} \G_{b}^{\ul} \h_{k}] & = G_{b}^{\ce} (G_{b}^{\ul})^{2} M, \\
\Exp \big[ |\v_{k}^{\herm} \G_{b}^{\ul} \h_{i}|^{2} \big] & = \Exp \big[ \| \G_{b}^{\ul} \v_{k} \|^{2} \big] \\
\nonumber & = \bigg( 1 + \frac{1}{\rho_{\bs} \tau} \bigg) (G_{b}^{\ce})^{2} (G_{b}^{\ul})^{4} M \\
& \phantom{=} \ + \frac{1}{\rho_{\bs} \tau^{2}} (G_{b}^{\ul})^{4} A_{k} \qquad \textrm{(}i \neq k\textrm{)}, \\
\nonumber \Exp \big[ |\v_{k}^{\herm} \G_{b}^{\ul} \h_{k}|^{2} \big] & = \bigg( M + 1 + \frac{1}{\rho_{\bs} \tau} \bigg) (G_{b}^{\ce})^{2} (G_{b}^{\ul})^{4} M \\
& \phantom{=} \ + \frac{1}{\rho_{\bs} \tau^{2}} (G_{b}^{\ul})^{4} A_{k}, \\
\nonumber \Exp [\v_{k}^{\herm} \C_{\d}^{\ul} \v_{k}] & = \bigg( 1 + \frac{1}{\rho_{\bs} \tau} \bigg) (G_{b}^{\ce})^{2} (G_{b}^{\ul})^{2} \tr (\C_{\d}^{\ul}) \\
& \phantom{=} \ + \frac{1}{\rho_{\bs} \tau^{2}} (G_{b}^{\ul})^{2} B_{k}^{\ul}.
\end{align}
This allows us to write the effective uplink SINDR with imperfect CSI and MRC as in \eqref{eq:gamma^ul_2} at the top of the next page. Finally, we substitute the resulting expression into \eqref{eq:R^ul} to obtain an uplink ergodic achievable sum rate.

\subsection{Downlink Ergodic Achievable Sum Rate with MRT} \label{sec:SR_imperf_DL}

Assuming that MRT is used at the BS, we have\footnote{We use $\mathrm{unvec} (\cdot)$ as the inverse of the $\mathrm{vec}(\cdot)$ operator.} $\W = \frac{1}{\sqrt{\delta}} \hat{\H}$, with $\delta \triangleq \Exp \big[ \| \hat{\h} \|^{2} \big]$ and $\hat{\H} \triangleq \mathrm{unvec} (\hat{\h}) \in \Compl^{M \times K}$, and the precoding vector corresponding to UE~$k$ is given by
\addtocounter{equation}{+1}
\begin{align}
\w_{k} & = \frac{1}{\sqrt{\delta} \tau} \bigg( \bar{\P}_{k}^{\tran} \G_{b}^{\ce} \bar{\P}^{*} \h + \frac{1}{\sqrt{\rho_{\bs}}} \bar{\P}_{k}^{\tran} \G_{b}^{\ce} \z^{\ce} + \frac{1}{\sqrt{\rho_{\bs}}} \bar{\P}_{k}^{\tran} \d^{\ce} \bigg).
\end{align}
Under the i.i.d. Rayleigh fading assumption, we have
\begin{align}
\nonumber \Exp [\W \W^{\herm}] & = \frac{1}{\delta} \bigg( K \bigg( 1 + \frac{1}{\rho_{\bs} \tau} \bigg) (G_{b}^{\ce})^{2} \I_{M} \\
& \phantom{=} \ + \frac{1}{\rho_{\bs} \tau^{2}} \sum_{k=1}^{K} \bar{\P}_{k}^{\tran} \C_{\d}^{\ce} \bar{\P}_{k}^{*} \bigg)
\end{align}
(with $\Diag \big( \Exp [\W \W^{\herm}] \big) = \frac{1}{M} \I_{M}$), and we can thus write $\G_{b}^{\dl}$ as a scaled identity matrix, i.e., $\G_{b}^{\dl} = G_{b}^{\dl} \I_{M}$ (see \eqref{eq:G_b_ul}). Then, we can express each expectation term in \eqref{eq:gamma^dl} as
\begin{align}
\Exp [\h_{k}^{\herm} \G_{b}^{\dl} \w_{k}] & = \frac{1}{\sqrt{\delta}} G_{b}^{\ce} G_{b}^{\dl} M, \\
\nonumber \Exp \big[ |\h_{k}^{\herm} \G_{b}^{\dl} \w_{i}|^{2} \big] & = \frac{1}{\delta} \bigg( \bigg( 1 + \frac{1}{\rho_{\bs} \tau} \bigg) (G_{b}^{\ce})^{2} (G_{b}^{\dl})^{2} M \\
& \phantom{=} \ + \frac{1}{\rho_{\bs} \tau^{2}} (G_{b}^{\dl})^{2} A_{i} \bigg) \qquad \textrm{(}i \neq k\textrm{)}
\end{align}
and
\begin{align}
\nonumber \Exp \big[ |\h_{k}^{\herm} \G_{b}^{\dl} \w_{k}|^{2} \big] & = \frac{1}{\delta} \bigg( \bigg( M + 1 + \frac{1}{\rho_{\bs} \tau} \bigg) (G_{b}^{\ce})^{2} (G_{b}^{\dl})^{2} M \\
& \phantom{=} \ + \frac{1}{\rho_{\bs} \tau^{2}} (G_{b}^{\dl})^{2} A_{i} \bigg), \\
\Exp [\h_{k}^{\herm} \C_{\d}^{\dl} \h_{k}] & = \tr (\C_{\d}^{\dl}).
\end{align}
This allows us to write the effective downlink SINDR with imperfect CSI and MRT as in \eqref{eq:gamma^dl_2} at the top of the page. Finally, we substitute the resulting expression into \eqref{eq:R^dl} to obtain a downlink ergodic achievable sum rate.

\begin{figure}[t!]
\centering
\includegraphics[scale=1]{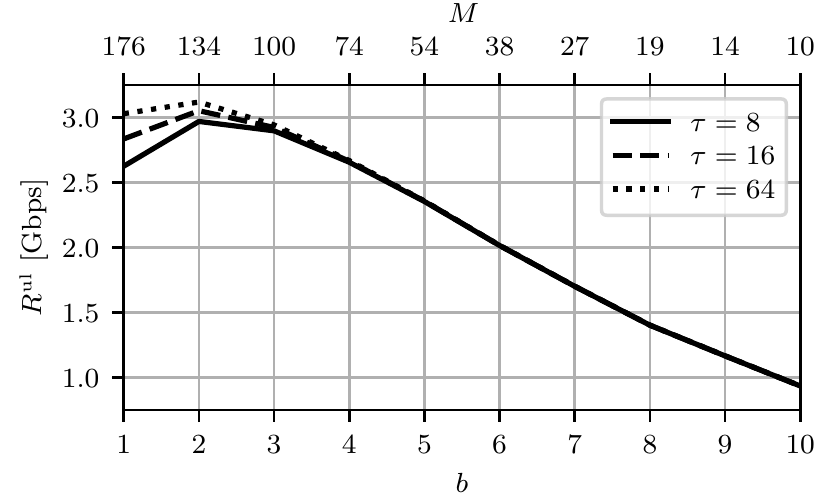}
\caption{\footnotesize Uplink ergodic achievable sum rate versus the resolution of the ADCs with $B = 0.1$~GHz and $P_{\hw} = 10 \big( P_{\rf} + 2 P_{\adc}(10~\textrm{bits},0.1~\textrm{GHz}) \big)$.} \label{fig:SR_ul_VS_b_0.1GHz} \vspace{-3mm}
\end{figure}

\section{Numerical Results and Discussion} \label{sec:NR}

We now evaluate the performance trade-off between the number of BS antennas, the resolution of the ADCs/DACs, and the bandwidth under hardware power consumption constraints. To do so, we use the uplink and downlink ergodic achievable sum rate expressions derived in Sections~\ref{sec:SR_imperf_UL} and~\ref{sec:SR_imperf_DL}, respectively, which consider imperfect CSI and MRC/MRT. Throughout we disregard the channel estimation overhead, which can be readily included by multiplying \eqref{eq:R^ul} and \eqref{eq:R^dl} by $\big( 1 - \frac{\tau}{T} \big)$ for a given coherence time $T$ (see, e.g., \cite[Sec.~4]{Bjo17}).

In the numerical results that follow, we consider $K = 8$ UEs and assume that $\P$ comprises the first $K$ columns of the $\tau$-dimensional discrete Fourier transform matrix. The SNR at the BS in the uplink (resp. at the UEs in the downlink) includes the transmit power at the UEs (resp. at the BS), the pathloss, and the thermal noise power. Hence, we have $\rho_{\bs} = \frac{P_{\ue} \beta_{k}}{\sigma_{\bs}^{2}}$, where $P_{\ue} = 20$~dBm is the transmit power at the UEs, $\beta_{k}$ represents the pathloss between the BS and UE~$k$, and $\sigma_{\bs}^{2}$ is the thermal noise power at the BS. Likewise, we have $\rho_{\ue} = \frac{P_{\bs} \beta_{k}}{\sigma_{\ue}^{2}}$, where $P_{\bs} = 30$~dBm is the transmit power at the BS and $\sigma_{\ue}^{2}$ is the thermal noise power at the UEs. The pathloss for each UE~$k$ is expressed as $\beta_{k} = d_{k}^{- \alpha}$, where $d_{k}$ is the distance between the BS and UE~$k$ and $\alpha = 4$ is the pathloss exponent. For simplicity, all the UEs are placed at the same distance from the BS, with $\{ d_{k} = 100~\textrm{m} \}_{k=1}^{K}$. The thermal noise power at both the BS and the UEs is given by $\sigma_{\bs}^{2} = \sigma_{\ue}^{2} = \nu - 174 + 10 \log_{10} (B)$ [dBm], where $\nu = 13$~dB is the noise figure; note that a larger bandwidth corresponds to an increased thermal noise power. Lastly, the covariance matrices of the quantization distortion $\C_{\d}^{\ul}$, $\C_{\d}^{\dl}$, and $\C_{\d}^{\ce}$ are computed via Monte Carlo simulations with $10^{6}$ independent channel realizations.

\begin{figure}[t!]
\centering
\includegraphics[scale=1]{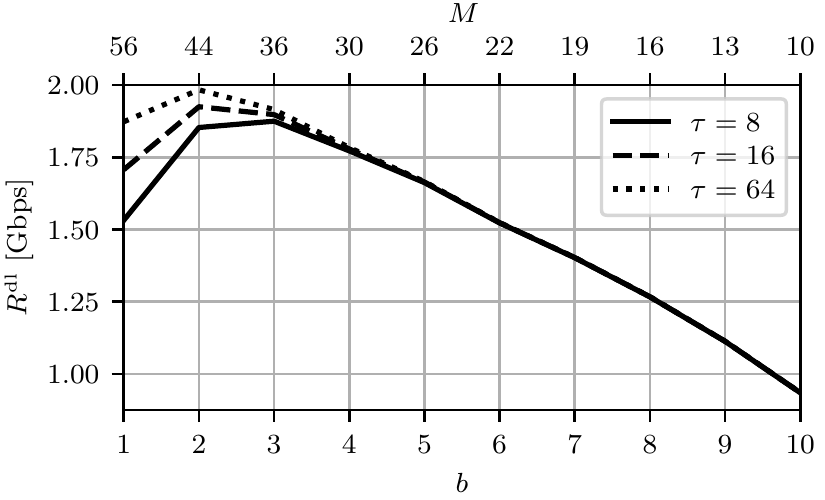}
\caption{\footnotesize Downlink ergodic achievable sum rate versus the resolution of the DACs with $B = 0.1$~GHz and $P_{\hw} = 10 \big( P_{\rf} + 2 P_{\dac}(10~\textrm{bits},0.1~\textrm{GHz}) \big)$.} \label{fig:SR_dl_VS_b_0.1GHz} \vspace{-3mm}
\end{figure}

\subsection{ADCs/DACs Setting} \label{sec:NR_quant}

We assume that the quantization thresholds and labels for each $b$-bit ADC/DAC are computed following the same procedure as in \cite{Jac17}. For the ADCs, the Lloyd-Max algorithm is applied to the Gaussian distribution with variance $\rho_{\ue} K + 1$, obtaining the thresholds $\setT_{b}$ and a set of labels $\tilde{\setL}_{b} \triangleq \{ \tilde{\ell}_{n} \}_{n=0}^{2^{b}-1}$.\footnote{This approach minimizes the mean squared error between the non-quantized and quantized signal.} Then, we obtain the labels $\setL_{b} = \zeta \tilde{\setL}_{b}$, with
\addtocounter{equation}{+1}
\begin{align} \label{eq:zeta_ul}
\zeta \triangleq \eta^{\ul} \bigg( \sum_{n=0}^{2^{b}-1} \tilde{\ell}_{n}^{2} \Big( \Phi \big( \eta^{-1} |t_{n+1}| \big) - \Phi \big( \eta^{-1} |t_{n}| \big) \Big) \bigg)^{-\frac{1}{2}}
\end{align}
with $\eta^{\ul} \triangleq \sqrt{\frac{\rho_{\ue} K + 1}{2}}$ and where $\Phi( \cdot )$ denotes the cumulative distribution function of the standard Gaussian distribution. Note that such a rescaling of the labels ensures that the variance of each entry of $\r^{\ul}$ in \eqref{eq:r^ul} and $\r^{\ce}$ in \eqref{eq:r^ce} is $\rho_{\ue} K + 1$. We follow a similar procedure for the DACs, where the Lloyd-Max algorithm is applied to the Gaussian distribution with variance $\frac{1}{M}$ and $\eta^{\ul}$ in \eqref{eq:zeta_ul} is replaced by $\eta^{\dl} \triangleq \frac{1}{\sqrt{2 M}}$.

To determine the number of BS antennas as a function of the resolution of the ADCs/DACs under a hardware power consumption constraint (see \eqref{eq:M}), we use the following power consumption model for the ADCs and DACs \cite{Cui05}:
\begin{align}
P_{\adc}(b,B) & = 3 V_{\textrm{dd}}^{2} L_{\textrm{min}} (2 B + f_{\textrm{cor}}) 10^{0.1525 b - 4.838}, \\
P_{\dac}(b,B) & = \frac{1}{2} V_{\textrm{dd}} I_{0} (2^{b} - 1) + b C_{\textrm{p}} (2 B + f_{\textrm{cor}}) V_{\textrm{dd}}^{2}
\end{align}
where $V_{\textrm{dd}} = 3$~V, $L_{\textrm{min}} = 0.5$~$\mu$m, $f_{\textrm{cor}} = 1$~MHz, $I_{0} = 10$~$\mu$A, and $C_{\textrm{p}} = 1$~pF are chosen as in \cite[App.~E]{Cui05}.

\begin{figure}[t!]
\centering
\includegraphics[scale=1]{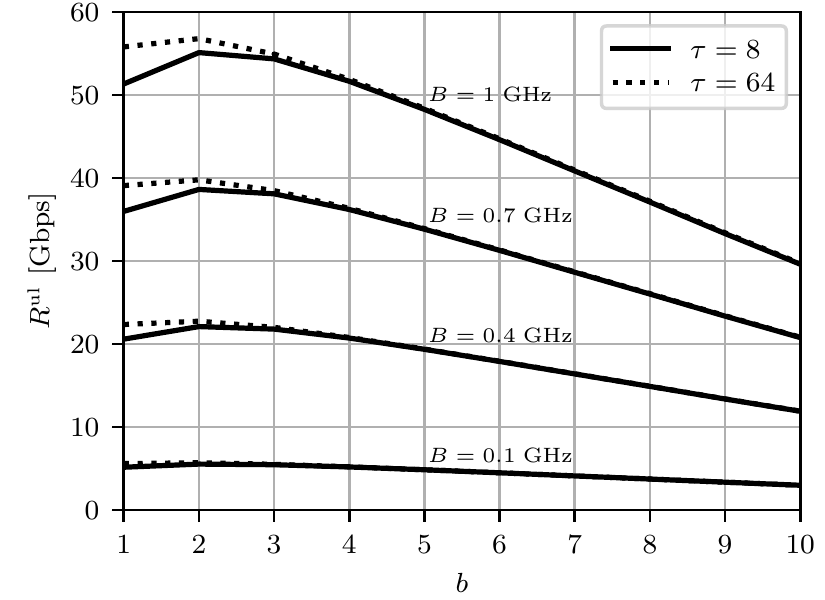}
\caption{\footnotesize Uplink ergodic achievable sum rate versus the resolution of the ADCs with $B \in \{0.1, 0.4, 0.7, 1\}$~GHz and $P_{\hw} = 10 \big( P_{\rf} + 2 P_{\adc}(10~\textrm{bits},1~\textrm{GHz}) \big)$.} \label{fig:SR_ul_VS_b_1GHz} \vspace{-3mm}
\end{figure}

\subsection{Performance Evaluation} \label{sec:NR_fig}

\begin{figure}[t!]
\centering
\includegraphics[scale=1]{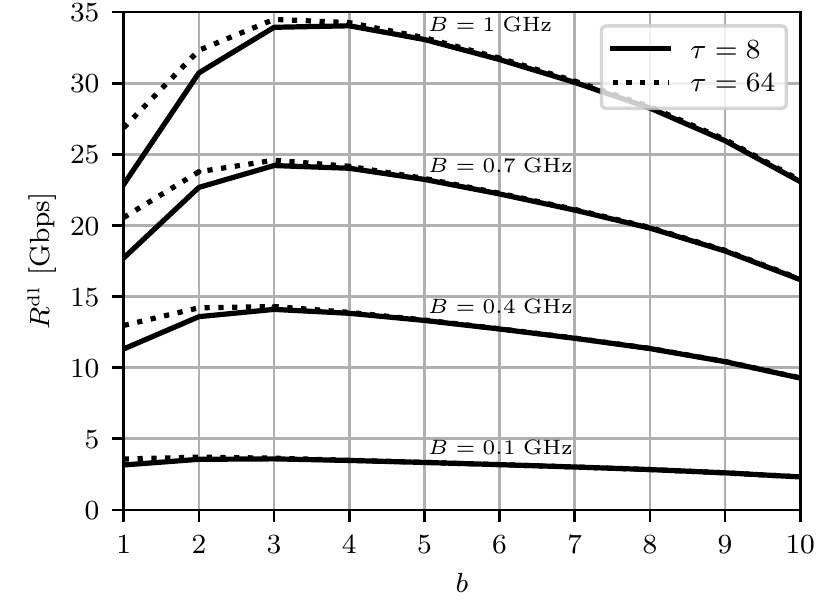}
\caption{\footnotesize Downlink ergodic achievable sum rate versus the resolution of the DACs with $B \in \{0.1, 0.4, 0.7, 1\}$~GHz and $P_{\hw} = 10 \big( P_{\rf} + 2 P_{\dac}(10~\textrm{bits},1~\textrm{GHz}) \big)$.} \label{fig:SR_dl_VS_b_1GHz} \vspace{-3mm}
\end{figure}

Let us assume $P_{\rf}=40$~mW in the uplink (as in \cite{Jos18}) and $P_{\rf}=10$~mW in the downlink. In Fig.~\ref{fig:SR_ul_VS_b_0.1GHz} and Fig.~\ref{fig:SR_dl_VS_b_0.1GHz}, we plot the uplink and downlink ergodic achievable sum rates versus the resolution of the ADCs and DACs, respectively, with $B = 0.1$~GHz and where $P_{\hw}$ is set such that it can supply $10$ BS antennas with $10$-bit ADCs/DACs at such bandwidth; note that the value of $M$ corresponding to each value of $b$ is detailed at the top of the figures. In general, lowering the resolution of the ADCs/DACs under a hardware power consumption constraint has a two-fold effect: on the one hand, the reduced~power consumption allows one to increase the number of BS antennas, which enhances the array gain; on the other hand, the quantization distortion becomes more significant during both the channel estimation and the uplink/downlink data transmission. In the uplink, the ergodic performance is maximized with $2$-bit ADCs for all the considered pilot lengths. Moreover, in the downlink, the best ergodic performance is achieved with $3$-bit DACs for $\tau = 8$ and with $2$-bit DACs for $\tau \geq 16$. In both the uplink and the downlink, the impact of imperfect CSI is more evident for lower resolutions and the extreme case of $1$-bit ADCs is nearly optimal in the uplink when accurate channel estimates are available. In Fig.~\ref{fig:SR_ul_VS_b_1GHz} and Fig.~\ref{fig:SR_dl_VS_b_1GHz}, we plot the uplink and downlink ergodic achievable sum rates versus the resolution of the ADCs and DACs, respectively, with different values of $B$ and where $P_{\hw}$ is set such that it can supply $10$ BS antennas with $10$-bit ADCs/DACs at $B = 1$~GHz. In the uplink, the ergodic performance is maximized with $2$-bit ADCs for all the considered bandwidths. In the downlink, the optimal number of resolution bits increases with the bandwidth.

\section{Conclusions} \label{sec:CON}

Considering a fully digital massive MIMO architecture with low-resolution ADCs/DACs at the BS, we derive tractable expressions for uplink and downlink ergodic achievable sum rates with imperfect CSI. These are used to analyzes the performance trade-off between the number of BS antennas, the resolution of the ADCs/DACs, and the bandwidth under a hardware power consumption constraint, where the relationship between these design parameters follows from a realistic model for the power consumption of the ADCs/DACs and the RF chains. In the evaluated scenarios, the most energy-efficient configurations feature $2$ resolution bits in the uplink and $2$ to $3$ resolution bits in the downlink. Future work will consider correlated channels as well as power consumption models reflecting the latest advancements in ADC/DAC technologies.

\bibliographystyle{IEEEtran}
\bibliography{IEEEabbr,refs}

\end{document}